\documentclass[a4paper,11pt]{article}
\usepackage{mathrsfs}
\usepackage{graphicx}

\usepackage[usenames]{color}

\usepackage{color}
\usepackage{colordvi}
\usepackage{amssymb,amsmath}
\usepackage{amsmath}
\usepackage{psfrag, graphicx}
\usepackage{picinpar}
\usepackage{graphics}
\usepackage{graphicx}
\usepackage{epsfig}
\usepackage{amssymb}

\newcommand{\beq}{\begin{equation}}
\newcommand{\eeq}{\end{equation}}

\def\bar#1{\begin{array}{#1}}
\def\ear{\end{array} }

\def\pmb#1{\setbox0=\hbox{$#1$}%
  \kern-.025em\copy0\kern-\wd0
  \kern.05em\copy0\kern-\wd0
  \kern-.025em\raise.0433em\box0}

\def\centerwmf#1#2#3{\vskip#2\relax\centerline{\hbox to#1{\special {wmf:#3 x=#1, y=#2}\hfil}}}

\def\bar{\overline}

\def\text{\rm}
\def\text{\mbox}

\begin {document}

\title{{\bf Scalar field cosmology in phase space}}
\author{{\bf Valerio Faraoni}\thanks{Physics Department 
and {\em STAR} Research Cluster,  
Bishop's University, 2600 College Street, Sherbrooke, 
Qu\'ebec, Canada J1M 1Z7} ~~and~~{\bf Charles S. 
Protheroe}\thanks{Physics Department, 
Bishop's University, 2600 College Street, Sherbrooke, 
Qu\'ebec, Canada J1M 1Z7} }
\date{}
\maketitle

\begin {abstract}
Using dynamical systems methods, we describe the  
evolution of a minimally 
coupled scalar field and a  
Friedmann-Lema\^{i}tre-Robertson-Walker universe in the 
context of general relativity, which is relevant for 
inflation and late-time quintessence eras.  Focussing  
on the spatially flat case, we examine the 
geometrical structure of the phase space, locate the 
equilibrium points of the system (de Sitter spaces with  
a constant  scalar field), study their stability through 
both 
a third-order perturbation analysis and Lyapunov 
functions, and discuss the late-time asymptotics. As we do 
not specify the scalar field's origin or its  potential, 
the results are independent of the high-energy 
model. 
\end {abstract}

\begin{quote}
\end{quote}

\section{Introduction}
\setcounter{equation}{0}
	
The standard cosmological model based on the spatially 
homogeneous and isotropic  
Friedmann-Lema\^{i}tre-Robertson-Walker (FLRW) metric is
very successful at describing many observations at  
different scales. It has become normal to include an 
inflationary epoch \cite{Guth} in this model during the 
early universe. 
Although there is no direct proof that inflation actually 
occurred, and other scenarios should still be considered, 
the 1992 discovery of temperature fluctuations in the 
cosmic microwave
background  by the {\em COBE} satellite  \cite{COBE} 
provided 
evidence  of a nearly scale-invariant spectrum of 
primordial density perturbations of the kind predicted by 
inflationary  scenarios. In addition, the study of these 
temperature fluctuations initiated by {\em COBE} ushered in 
an era
of ``precision cosmology'' continued with later cosmic 
microwave background experiments, most notably {\em WMAP} 
and {\em PLANCK} \cite{WMAP, Planck}. Most models of early 
universe  inflation are based on scalar fields, and those 
based on quadratic quantum corrections to the 
Einstein-Hilbert action 
(``Starobinsky inflation'' \cite{Starobinskyinflation}) 
can be reduced to the study of scalar field degrees of 
freedom \cite{SotiriouFaraonireview}. 

A second revolution in cosmology occurred in 1998 with 
the discovery, obtained by studying type Ia supernovae, 
that the current expansion of the universe is accelerated 
\cite{SN}. In the context of general relativity, on which 
the standard $\Lambda$-cold dark matter model is based, 
this acceleration can only be explained with a cosmological 
constant $\Lambda$ of extremely fine-tuned, but 
non-vanishing, magnitude, or with a very exotic fluid 
having pressure $P$ and density $\rho$ related by the 
equation of state~ $P\approx-\rho$, and dubbed ``dark 
energy''. Most models of dark energy are 
based on a scalar field $\phi$ (also known as 
``quintessence'')  rolling in a flat section of 
its potential $V(\phi)$. Alternative  scenarios, seeking to 
replace the Einstein-Hilbert action (``$f(R)$'' or 
``modified''  gravity \cite{SotiriouFaraonireview}), can 
again be reduced to the dynamics of a scalar field degree  
of freedom.

Both inflation and quintessence models mandate a general 
understanding of scalar field dynamics in 
general-relativistic cosmology.  Furthermore, a scalar 
field provides the simplest field  theory of matter, and 
although no fundamental classical scalar field has been 
discovered in nature so far (except possibly 
for quintessence),  they do provide a toy model useful for 
understanding many basic theoretical features of more 
realistic field 
theories, without the extra details and complications. As 
such, scalar field theory also constitutes  an excellent 
pedagogical tool 
used in most relativity textbooks. 

In this paper we approach the spatially homogeneous and 
isotropic cosmology of scalar fields minimally coupled 
to gravity from the phase space point of view. Although 
dynamical system methods have been widely used in cosmology 
since the 1960s \cite{cosmodynsyst} and this type 
of analysis has been performed for non-minimally coupled 
scalar fields \cite{phasespaceNMC} and general 
scalar-tensor or $f(R)$ gravity \cite{VFAnnPhys, 
DeSouzaFaraoni}, we could not find in the literature a 
complete and self-contained analysis for the simpler  
case 
of relativity with a minimally coupled scalar field,  
apart from specific scenarios corresponding to particular 
choices of the scalar field potential $V(\phi)$, which 
abound in the literature ({\em e.g.},  
\cite{specificscenarios}-\cite{LiddleLythbook}, see also 
\cite{extra}---the literature 
on specific scenarios is very 
large and here we limit ourselves to quote the papers whose 
approach is closest to the general one that we 
adopt).\footnote{Ref.~\cite{Foster} studies the 
minimally coupled case of interest here and presents some 
of the features of scalar field cosmology derived in the 
following, but its main  interest is in initial  
singularities and particular classes of potentials.} By  
contrast, here  we do not 
commit to any particular scenario,  and at most, we make 
general assumptions on properties of 
the potential (such as boundedness or monotonicity), 
refraining from choosing specific forms of the function 
$V(\phi)$.   Given that there is no preferred scenario of 
inflation or quintessence, general considerations are 
valuable.

Since the relevant equations, which reduce 
to ordinary differential equations (ODEs) in this case, are 
still non-linear and not amenable to exact solution, the 
phase space view becomes important in gaining a qualitative 
understanding of the solutions without actually solving the 
field equations. It is generally believed that in order to 
say anything about the phase space and the qualitative 
behaviour of the solutions of the  equations, 
one must first fully specify  the scenario of inflation
or quintessence being studied. While this is certainly true 
if one wants a complete qualitative picture of the 
dynamics, many aspects of the phase space portrait are 
common to most, if not all, scenarios and the study of 
these aspects, without committing to any particular
scenario or potential $V(\phi)$, is a necessary preliminary  
for more detailed 
analyses of specific models. The purpose of this paper is 
to discuss these general features, specifically the 
geometry of the phase space, the existence, nature, and 
stability of the fixed points, and the late-time behaviour
of the solutions, without specifying the form of the scalar 
field potential energy density, and instead 
making some generic assumptions on its behaviour 
(boundedness, presence of asymptotes, {\em etc.}).

\section{Background}
\setcounter{equation}{0}

We consider a scalar field minimally coupled to the 
spacetime curvature as the only source of gravity in the 
Einstein field  equations.  This assumption is fully 
justified in 
inflationary scenarios of the early universe 
\cite{LiddleLythbook}, and only  approximately justified in 
quintessence models of the late  universe 
\cite{AmendolaTsujikawabook}. In the latter case, 
scalar field dark energy is present along with a dust 
fluid, which combine to determine the dynamics of the 
universe. However, observations suggest that dark energy 
comes to dominate the dynamics very quickly,
starting from redshifts $z\sim0.5$, thus we can once again 
neglect the dust fluid and other forms of energy in the 
late regimes. In short, there is plenty of motivation to 
study scalar field cosmology.

The Lagrangian density of a scalar field $\phi$ minimally 
coupled to the spacetime curvature is\footnote{We follow 
the notations of Ref.~\cite{Wald}.}
\begin{equation}
 \mathcal{L}^{(\phi)}=-\frac{1}{2} 
\, \nabla^{\alpha}\phi\nabla_{\alpha} \phi - V(\phi) 
\,, \label{phiLagrangian}
\end{equation}
where $V(\phi)$ is the scalar field potential. The action 
for gravity and the scalar field is
\begin{equation}
S=\int d^{4}x\sqrt{-g}\left[\frac{R}{16\pi G} 
+\mathcal{L}^{(\phi)}(\phi,\, g_{\mu\nu})\right] 
 \equiv S_{(g)}+S_{(\phi)} \,,\label{eq:actionS}
\end{equation}
where $g_{\mu\nu}$ is the spacetime metric, $g$ is its 
determinant, and $R$ is its Ricci scalar. The action 
(\ref{eq:actionS}) is also the action for  general 
scalar-tensor gravity 
{\em in vacuo}, after performing a conformal 
transformation to the Einstein frame ({\em e.g.}, 
\cite{mybook}).  The variation of 
the scalar field action $S_{(\phi)}=\int 
d^{4}x\,\sqrt{-g} \, \mathcal{L}^{(\phi)}$ gives the
stress-energy tensor
\begin{equation}
T_{\mu\nu}^{(\phi)}= 
\frac{-2}{\sqrt{-g}}\frac{\delta 
\mathcal{L}^{(\phi)}}{\delta g^{\mu\nu}} 
=\nabla_{\mu}\phi\nabla_{\nu}\phi-\frac{1}{2} 
\, g_{\mu\nu} \nabla^{\alpha}\phi\nabla_{\alpha} \phi 
-V(\phi) \, g_{\mu\nu} \,.
\label{eq:scalarfieldstress-energytensor} 
\end{equation}
A spatially homogeneous and isotropic universe is 
described by the FLRW line element
\begin{equation}
ds^{2}=-dt^{2}+a^2(t)\left[ 
\frac{dr^{2}}{1-kr^{2}}+r^{2}
\left(d\theta^{2}+\sin^2 \theta 
\, d\varphi^{2}\right)\right]
\end{equation}
in comoving coordinates $\left( t,r,\theta,\varphi 
\right)$, where $a(t)$ is the scale factor and $k$ is the 
curvature index. The Einstein equations
\begin{equation}
R_{\mu\nu}-\frac{1}{2} \, g_{\mu\nu} R = 8\pi 
G \, T_{\mu\nu}\label{eq:EinsteinEqs}
\end{equation}
(where $R_{\mu\nu}$ is the Ricci tensor and $R\equiv 
{R^{\mu}}_{\mu}$) reduce to ODEs 
for the scale factor and matter degrees of freedom. 
It is customary to approximate the matter content of the 
universe with a single perfect fluid with four-velocity
$u^{\mu}=\delta^{0\mu}$ in comoving coordinates,  energy 
density $\rho$, pressure $P$, and energy-momentum tensor
\begin{equation}
T_{\mu\nu}= 
\left( P+\rho 
\right)u_{\mu}u_{\nu}+Pg_{\mu\nu} \,.
\label{eq:fluidstress-energytensor} 
\end{equation}
The pressure and energy density are usually related by a 
barotropic equation of state $P=P(\rho)$, often 
of the form $P=w\rho$ 
where the constant $w$ is called the ``equation of state 
parameter''. The Einstein equations (\ref{eq:EinsteinEqs}) 
in the presence of a single perfect
fluid reduce to 
\begin{eqnarray}
\frac{\ddot{a}}{a} =-\frac{4 \pi G}{3}  
\left(\rho+3P\right) \,,
\label{eq:fluidacceleration}\\
\nonumber\\
\frac{\dot{a}^{2}}{a^{2}}=\frac{8\pi 
G}{3}\rho-\frac{k}{a^{2}} \,,\label{eq:fluidhamilton}\\
\nonumber\\
\dot{\rho}+3 \, \frac{\dot{a}}{a}\left(P+\rho\right)=0 
\,, \label{eq:fluidKG}
\end{eqnarray}
where an overdot denotes differentiation with respect to 
the comoving time $t$. Eqs. (\ref{eq:fluidacceleration}) 
and (\ref{eq:fluidhamilton}) are called the acceleration 
equation and the Hamiltonian constraint, respectively, and 
the Klein-Gordon equation (\ref{eq:fluidKG}) is nothing 
but  the covariant conservation equation 
$\nabla^{\nu}T_{\mu\nu}=0$ (when $\phi \neq $const.). The 
Klein-Gordon equation is not independent of eqs. 
(\ref{eq:fluidacceleration}) and (\ref{eq:fluidhamilton}) 
and can be derived from them. Excellent 
pedagogical analyses of the phase space of a FLRW universe 
coupled to a perfect  fluid are available in the literature 
\cite{FluidPhaseSpace}.

In a FLRW universe, a gravitating scalar field must   
necessarily depend only on the comoving time,  
$\phi=\phi(t)$, in order to respect the spacetime 
symmetries. Therefore, its gradient $\nabla_{\mu}\phi$
is timelike (or null but trivial if $\phi=$const.). In 
regions where $\nabla^{\alpha}\phi\nabla_{\alpha}\phi<0$, 
we can introduce the four-vector
\begin{equation}
u_{\mu}=\frac{\nabla_{\mu}\phi}{\sqrt{ 
\left|\nabla^{\alpha}\phi\nabla_{\alpha}\phi\right|}} \,,
\end{equation}
with $u_{\mu}u^{\mu}=-1$, and the scalar field is 
equivalent to a perfect fluid with stress-energy tensor of 
the form (\ref{eq:fluidstress-energytensor})
and energy density and pressure \cite{Madsen}
\begin{eqnarray}
\rho &=& \frac{\dot{\phi}^{2}}{2}+V(\phi) \,,\\
\nonumber\\
P &=& \frac{\dot{\phi}^{2}}{2}-V(\phi) \,.
\end{eqnarray}
One can define the effective equation of state parameter
\begin{equation}
w(\phi,\dot{\phi})\equiv\frac{P}{\rho} 
=\frac{\dot{\phi}^{2} - 2V(\phi)}{\dot{\phi}^{2}+2V(\phi)} 
\,.
\end{equation}
The Einstein-Friedmann equations 
(\ref{eq:fluidacceleration})--(\ref{eq:fluidKG}) become
\begin{eqnarray}
\frac{\ddot{a}}{a}=-\frac{8\pi 
G}{3}\left(\dot{\phi}^{2}-V(\phi)\right) 
\,,\label{eq:accelerationeq} \\
\nonumber\\
\left(\dot{\frac{a}{a}}\right)^{2}
=\frac{8\pi G}{3}\left(\frac{\dot{\phi}^2}{2} 
+V(\phi)\right)-\frac{k}{a^{2}} \,,
\label{eq:hamilconstraint} \\ 
\ddot{\phi}+3 \, \frac{\dot{a}}{a}\dot{\phi} 
+\frac{dV}{d\phi}=0 \,.
\label{eq:KleinGordon}
\end{eqnarray}
In the following it will be useful to rewrite these 
equations in terms of the Hubble parameter 
$H\equiv\dot{a}/a$ as
\begin{eqnarray}
\dot{H}&=&-H^{2}-\frac{8\pi G}{3}\left(\dot{\phi}^{2}
-V(\phi)\right)+\frac{k}{a^{2}}=-4\pi G\dot{\phi}^{2}
+\frac{k}{a^{2}} \,, \label{eq:FriedH-dot} \\
\nonumber\\
H^{2}&=&\frac{8\pi 
G}{3}\left(\frac{\dot{\phi}^{2}}{2}+V(\phi)\right) 
- \frac{k}{a^2}  \, ,\label{eq:FriedH^2} \\
\nonumber\\
\ddot{\phi} &+ & 3H\dot{\phi}+V^{'}=0 \,,\label{eq:FriedKG}
\end{eqnarray}
where a prime denotes dfferentiation with respect to 
$\phi$. These equations can also be derived from 
an effective Lagrangian or Hamiltonian \cite{Noether}.

The equations of scalar field cosmology are non-linear and 
few exact solutions are known for particular choices of the 
potential $V(\phi)$. We would like to discuss the dynamics 
of the variables $a(t)$ and $\phi(t)$ in as much 
depth  as  possible without choosing a specific form of 
$V(\phi)$. Before we begin, let us note that

\begin{itemize}

\item for $V(\phi)=0$ the scalar field is equivalent to a 
fluid with stiff equation of state $P=\rho$, which does not 
seem to be very relevant for inflation and late-time 
acceleration  
(although it is relevant for matter  at nuclear densities 
in the core of neutron stars, and possibly near the 
Big Bang singularity \cite{Foster}).

\item For $V(\phi)=V_0=$~const. the potential reduces to 
a pure cosmological constant $\Lambda$. The scalar field 
stress-energy tensor (\ref{eq:scalarfieldstress-energytensor})
reduces to 
\begin{equation}
T_{\mu\nu}=-\frac{\Lambda}{8\pi G} \, g_{\mu\nu} 
-\partial_{\mu}\phi\partial_{\nu}\phi-\frac{1}{2}\,  
g_{\mu\nu} \, \partial^{\alpha}\phi\partial_{\alpha}\phi 
\,,
\end{equation}
with $\Lambda=8\pi GV_{0}$. Further, for 
$\phi=\phi_0=$const., one recovers the stress-energy tensor 
of a pure cosmological constant.
 \end{itemize}

\section{Phase space}
\setcounter{equation}{0}

Eqs. (\ref{eq:accelerationeq}) and (\ref{eq:KleinGordon}) 
describe the evolution of $a(t)$ and $\phi(t)$ (remember 
that there are only two independent equations in the set 
(\ref{eq:accelerationeq})--(\ref{eq:KleinGordon}) if $\phi$ 
is not constant). Eq. (\ref{eq:hamilconstraint}) is a first 
order constraint (contrary to eqs. (\ref{eq:accelerationeq})
and (\ref{eq:KleinGordon}) which are of second order). The 
phase space is, therefore, a 4-dimensional space 
$\left(a,\dot{a},\phi,\dot{\phi}\right)$, but the 
Hamiltonian constraint (\ref{eq:hamilconstraint}) forces
the orbits of the solutions to live on a 3-dimensional 
hypersurface, introducing a relation between the four 
variables. For example, one can use the constraint to 
express $\dot{\phi}$ in terms of the other three variables, 
$\dot{\phi}=\dot{\phi}\left( a,\dot{a},\phi \right)$. 

For particular choices of the scalar field potential, and 
especially for $k\neq0$, one can change variables to 
functions of $a,\dot{a},\phi,\dot{\phi}$ which can lead to 
exact solutions or to simpler calculations. In general, 
however, these new variables do not have an immediate or 
clear physical meaning and are to be regarded as a 
mere mathematical trick to perform calculations. Often the 
results of these calculations cannot be translated 
 explicitly or easily in terms of the  
variables $\left(a,\dot{a},\phi,\dot{\phi}\right)$.
However, current observations seem to indicate that we 
live in a spatially flat $(k=0)$ universe, which is much 
simpler to analyze than the $k\neq0$ case. This is the 
situation  that we consider in the following.

\section{Spatially flat FLRW scalar field cosmologies}
\setcounter{equation}{0}

The description of the phase space greatly simplifies for 
$k=0$ as, in this case, the scale factor $a(t)$ appears in 
the dynamical equations  only through the combination 
$\dot{a}/a\equiv H$, the Hubble parameter, which is a 
physical observable obtained by fitting 
theoretical models to cosmological data. Since $\phi$ is 
the only matter field in the theory, it is natural 
from the field theory point of view to choose 
it as another dynamical variable.\footnote{An  
astronomer would instead choose the density of the matter 
field $\Omega_{\phi}$ (in units 
of the critical density) as another variable.} By choosing 
$H$ and $\phi$ as dynamical variables, the phase space 
reduces to the 
3-dimensional space $\left(H,\phi,\dot{\phi}\right)$, but 
the orbits of the solutions of eqs. 
(\ref{eq:FriedH-dot})--(\ref{eq:FriedKG}) with $k=0$ are 
forced to move on a 2-dimensional subset of the phase 
space by  the Hamiltonian constraint (\ref{eq:FriedH^2}). 

Let us examine the structure of the ``energy surface'' on 
which the orbits are forced to move. We choose to eliminate 
$\dot{\phi}$ by expressing it in terms of the other 
variables $(H,\phi)$ in eq. (\ref{eq:FriedH^2}) with $k=0$, 
which can then be viewed formally as a quadratic algebraic 
equation for $\dot{\phi}$ and solved, obtaining
\begin{equation}
\dot{\phi}=\pm\sqrt{\frac{3H^2}{4\pi G} -2V(\phi)} \,.
\label{eq:phasespaceconstraint}
\end{equation}
For certain choices of the potential $V(\phi)$, an 
arbitrary choice of values of the pair $\left( H,\phi 
\right)$ could make the argument of the square root on the 
right hand side negative. Therefore, in general, there can 
be a region of the phase space forbidden to the orbits of the
dynamical solutions,
\begin{equation}
\mathcal{F}\equiv\left\{  
\left(H,\phi,\dot{\phi}\right):\,\,3H^{2}<8\pi G 
\, V(\phi)\right\} 
\end{equation}
(``forbidden region''). This region may or may not exist 
depending on the form of $V(\phi)$.

\begin{figure}[t]
\centering
\includegraphics[width=0.67\textwidth]{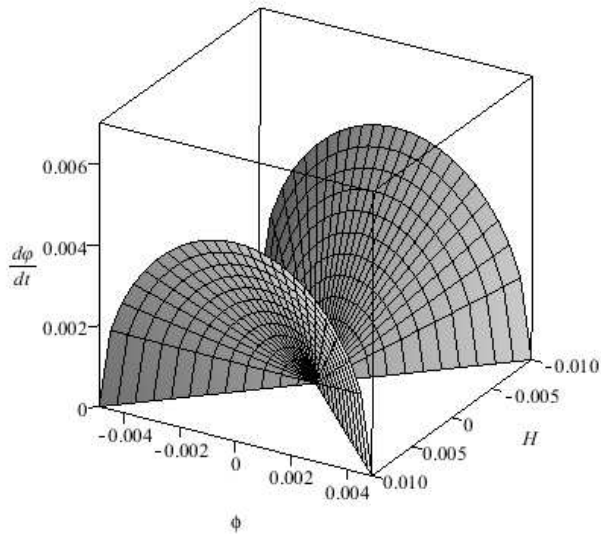}
\caption{The upper sheet corresponding to the positive sign 
in eq.~(\ref{eq:phasespaceconstraint}), for the quadratic 
potential $V(\phi)=m^2\phi^2/2$ (in arbitrary units).} 
\label{fig:fig1}
\end{figure}

\begin{figure}[t]
\centering
\includegraphics[width=0.67\textwidth]{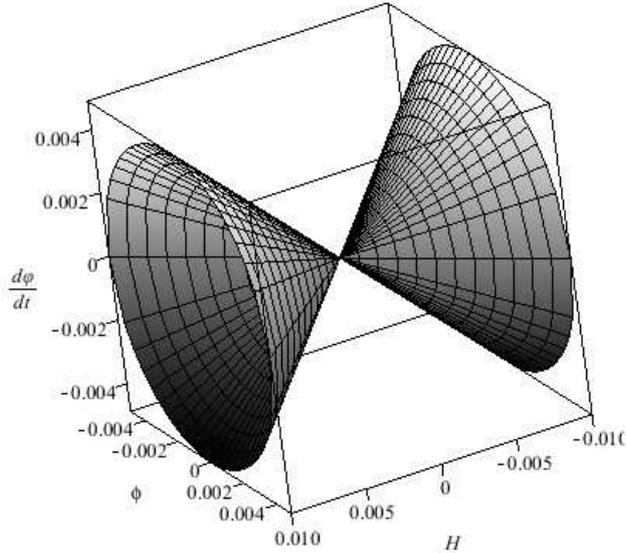}
\caption{The surface describing the Hamiltonian constraint 
(\ref{eq:phasespaceconstraint}) for the quadratic 
potential 
$V(\phi)=m^2\phi^2/2$ (in arbitrary units). The upper and 
lower sheets join at the $\dot{\phi}=0$ plane to form a 
cone.} \label{fig:fig2}
\end{figure}

There are two portions of the phase space region 
accessible to the dynamics (the ``energy surface'' 
corresponding to vanishing effective Hamiltonian 
\cite{Noether}), corresponding to the two signs of the 
right hand side of eq. (\ref{eq:phasespaceconstraint}).
These sets are symmetric with respect to the $\dot{\phi}=0$ 
plane of the $\left(H,\phi,\dot{\phi}\right)$ space.  We 
call 
these two subsets of the energy surface ``upper sheet'' and 
``lower sheet'', corresponding to the positive and negative 
sign, respectively. In the
upper sheet $\phi$ is always increasing 
$\left(\dot{\phi}>0\right)$, while on the lower sheet 
$\phi$ is always decreasing $\left(\dot{\phi}<0\right)$.
The two sheets are either disconnected, or always join on 
the plane
$\dot{\phi}=0$, on the boundary of the forbidden region
\begin{equation}
\mathcal{B}\equiv \partial\mathcal{F}=\left\{ 
\left(H,\phi,\dot{\phi}\right):\,\,\dot{\phi}=0
\Leftrightarrow3H^{2}=8\pi GV(\phi)\right\} \,.
\end{equation}
Figs.~\ref{fig:fig1} and \ref{fig:fig2} show the upper and 
lower  sheet for   the example  potential 
$V(\phi) =m^2\phi^2/2$. The dynamics of the spatially 
curved ($k \neq 0$) scalar 
field universe are confined to either side of the ``energy 
surface''  corresponding to $k=0$ in the phase space---this 
fact can be deduced by reducing the Hamiltonian 
constraint to 
\begin{equation}
\dot{\phi}= \pm \sqrt{\frac{3H^2}{4\pi G} 
-2V(\phi) + \frac{3k}{4\pi G a^2}} \,.
\end{equation}
Trajectories corresponding to $k > 0$ would exist 
above the $k = 0$ upper sheet ({\em i.e.}, for larger 
values of $\dot{\phi}$ than those corresponding to the 
$k=0$ upper sheet)  and below the $k=0$ lower sheet ({\em 
i.e.}, for lower values of $\dot{\phi}$). Trajectories 
corresponding to $k < 0$ would 
exist between  each $k = 0$  sheet ({\em i.e.}, for values 
of $\dot{\phi}$ comprised between those given by 
eq.~(\ref{eq:phasespaceconstraint})). This property was 
realized in  Ref.~\cite{Belinskyetal} for the specific  
inflationary potential $V=m^2\phi^2/2$, but the conclusion 
is general. Trajectories in a region corresponding to $k>0$ 
cannot  cross the $k=0$ sheet and move to regions 
corresponding to $k<0$, and {\em vice-versa}. Such 
dynamical transitions betwen  different topologies 
of the universe are forbidden (the topology of 
spacetime is not ruled by the dynamics).


Finally, the lower dimension of the ``energy 
surface'' leads one to believe that chaos is impossible in 
the dynamical system under study. This statement is not 
trivial given that the standard results on the absence of 
chaos in a two-dimensional phase space are proven for a 
plane, not for a curved surface or for a subset of a 
higher-dimensional phase space obtained by gluing two 
2-dimensional sheets \cite{Lyapunov}. However, it is not 
 difficult to reduce this situation to the standard 
case, as has been shown for scalar-tensor gravity in 
\cite{FaraoniJensenTheuerkauf}. The theory of a minimally 
coupled scalar field in Einstein gravity is contained in 
this reference as a special case.

\section{Equilibrium points}
\setcounter{equation}{0}

Having chosen $ H$ and $\phi$ as dynamical 
variables, the equilibrium points of the dynamical system 
(when they exist) are, by definition, of the form 
$\left(\dot{H},\dot{\phi}\right)\equiv(0,0)$  and 
$\left(\ddot{H},\ddot{\phi}\right)\equiv(0,0)$,
or $\left(H,\phi\right)=\left(H_{0},\phi_{0}\right) 
=\left(\mbox{const.} , \mbox{const.} \right)$, and they 
must all lie in the 
$\dot{\phi}=0$ plane, and therefore, on the boundary 
${\cal B}$ of the 
forbidden region (if this region exists). These 
equilibrium points are de  Sitter spaces with a  constant 
scalar field. When they  exist, they are the {\em only} 
de Sitter spaces possible in this theory. In fact, eq. 
(\ref{eq:FriedH-dot}) with $k=0$ reduces to $\dot{H}=-4\pi 
G\dot{\phi}^{2}$, and a de Sitter space with $H=$const.  
necessarily has $\phi=$const. as well.\footnote{By 
contrast, with non-minimally coupled scalar fields, de 
Sitter spaces with a non-constant scalar field are 
possible \cite{phasespaceNMC}.} A degenerate case is 
$H_{0}=0$, which corresponds to Minkowski space.
de Sitter spaces are important in cosmology because they 
are usually attractors in inflation and quintessence 
models \cite{LiddleLythbook, AmendolaTsujikawabook}. For 
$\phi=$const., $\mathcal{L}^{(\phi)}$ and 
$T_{\mu\nu}^{(\phi)}$ reduce to 
$\mathcal{L}^{(\phi)}=-V_{0}\equiv-V(\phi_{0})$
and $T_{\mu\nu}^{(\phi)}=-V_0 \, g_{\mu\nu}$, {\em i.e.}, 
to a pure cosmological constant term with 
$\Lambda=8\pi GV_{0}$.

The necessary and sufficient conditions for the existence 
of de Sitter fixed points are easily obtained from eqs. 
(\ref{eq:FriedH-dot})--(\ref{eq:FriedKG}) with $k=0$:
\begin{eqnarray}
H_0^2 &=& \frac{8\pi G}{3} \, V_{0} \,,
\label{eq:fixdptsH2}\\
&&\nonumber\\
 V_0^{'} &=& 0 \,,\label{eq:fixdptsV'}
\end{eqnarray} 
which obviously require $V_{0}\geq0$  (Minkowski space is 
obtained for $V_0=0$). Eq. (\ref{eq:fixdptsV'}) expresses 
the condition that $V(\phi)$ has an extremum or a point 
with horizontal tangent at $\phi_{0}$.

\begin{figure}[t]
\centering
\includegraphics[width=1.00\textwidth]{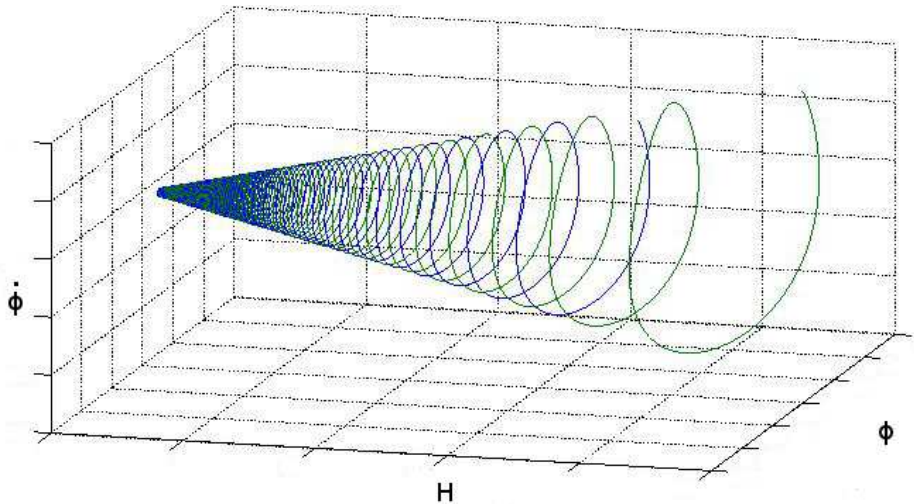}
\caption{Trajectories converging to a Minkowski fixed 
point $\left( H, \phi, \dot{\phi} \right)=\left( 0,0,0 
\right)$ for the previous example $V(\phi)=m^2\phi^2/2$.} 
\label{fig:fig3} \end{figure}

Fig.~\ref{fig:fig3} shows two trajectories, corresponding 
to different initial conditions, converging to a Minkowski 
space attractor point for the example of the 
$V(\phi)=m^2\phi^2/2$ potential.

Attractors (or repellors) could exist as an asymptotic 
limit in $a$ or $\phi$. To check for these we must search 
for fixed points with infinite values of the variables.

\subsection{Fixed points at infinity with a Poincar\'e  
projection}

Fixed points at infinity can be found by adopting polar 
coordinates $\left( r,\theta\right)$ with
\begin{equation}
H=r\cos\theta \, , \;\;\; \; \phi=r\sin\theta \, ,
\end{equation}
and the standard Poincar\'e transformation 
$r\rightarrow\bar{r}$ with
\begin{equation}
r\equiv\frac{\sqrt{\bar{r}}}{1-\bar{r}} 
\,,
\label{eq:Poincaretransf}
\end{equation}
which maps infinity onto the circle of radius $\bar{r}=1$. 
Since
\begin{equation}
\dot{H}=\frac{\left(1+\bar{r}\right)}{2\sqrt{\bar{r}} 
\left(1-\bar{r}\right)^2} 
\, \dot{\bar{r}}\cos\theta
-\frac{\sqrt{\bar{r}}}{1-\bar{r}} \, 
\dot{\theta}\sin\theta \,,
\end{equation}
\begin{equation}
\dot{\phi}=\frac{\left(1+\bar{r}\right)}{2\sqrt{\bar{r}} 
\left(1-\bar{r}\right)^2} \, 
\dot{\bar{r}}\sin\theta+
\frac{\sqrt{\bar{r}}}{1-\bar{r}} \, \dot{\theta}\cos\theta 
\,,
\end{equation}
fixed points  
$\left(H,\phi\right)=\left(\mbox{const.} , 
\mbox{const.} \right)$ correspond 
to $  \left(\bar{r},\theta\right)=\left( 
\mbox{const.}, \mbox{const.} \right)$
thanks to the linear independence of the sine and cosine 
functions. The 
dynamical system (\ref{eq:FriedH-dot})--(\ref{eq:FriedKG})  
becomes
\begin{eqnarray}
&&\frac{\bar{r}}{\left(1-\bar{r}\right)^{2}} 
\:\cos^{2}\theta  \nonumber\\
&&\nonumber\\
&& =   \frac{4\pi 
G}{3}\left[\left(\frac{\dot{\bar{r}} 
\left(1+\bar{r}\right)}{2\sqrt{\bar{r}} 
\left(1-\bar{r}\right)^{2}}
\:\sin\theta+\frac{\sqrt{\bar{r}}}{1-\bar{r}}
\:\dot{\theta}\cos\theta\right)^{2}+2V\right] 
\, , \label{eq:poincareH2}\\
&&\nonumber\\
&&\frac{\dot{\bar{r}}\left(1+\bar{r}\right)}{2\sqrt{\bar{r}} 
\left(1-\bar{r}\right)^{2}}\:\cos\theta 
- \, \frac{\sqrt{\bar{r}}}{1-\bar{r}} 
\:\dot{\theta}\sin\theta \nonumber\\
&&\nonumber\\
&&=  -4\pi 
G\left[\frac{\dot{\bar{r}} 
\left(1+\bar{r}\right)}{2\sqrt{\bar{r}}\left(1-\bar{r} 
\right)^{2}}\:\sin\theta+ \frac{\sqrt{\bar{r}}}{1-\bar{r}} 
\:\dot{\theta}\cos\theta\right]^{2} 
\,,\label{eq:poincareHdot}
\end{eqnarray}
\begin{eqnarray}
&&\left[\frac{\ddot{\bar{r}}+\ddot{\bar{r}}\bar{r}
+\dot{\bar{r}}^{2}}{2\sqrt{\bar{r}}\left(1-\bar{r}\right)^{2}}
- \, \frac{\dot{\bar{r}}^{2}+\dot{\bar{r}}^{2}\bar{r}}{ 
4\bar{r}^{3/2}\left(1-\bar{r}\right)^{2}} 
+ \, \frac{\dot{\bar{r}}^{2} 
+\dot{\bar{r}}^{2}\bar{r}}{4\sqrt{ 
\bar{r}}\left(1-\bar{r}\right)^{3}}\right] 
\sin\theta \nonumber\\
&&\nonumber\\
&& + \, \frac{\dot{\bar{r}}+\dot{\bar{r}}\bar{r}}{\sqrt{ 
\bar{r}}\left(1-\bar{r}\right)^{2}}\:\dot{\theta}\cos\theta 
+ \, \frac{\sqrt{\bar{r}}}{1-\bar{r}}\left(\ddot{\theta} 
\cos\theta-\dot{\theta}^{2}\sin\theta\right) \nonumber\\
&&\nonumber\\
&&=-3 \left[\frac{\bar{r}}{\left(1-\bar{r}\right)^{2}} 
\:\dot{\theta}\cos^{2}\theta+ \, \frac{\dot{\bar{r}} 
\left(1+\bar{r}\right)}{2\left(1-\bar{r}\right)^{3}} 
\:\sin\theta\right]  - V^{'} 
\,.\label{eq:poincareKG}
\end{eqnarray}
Setting 
$\left(\dot{H},\dot{\phi}\right)=\left(0,0\right)$
and $\left(\ddot{H},\ddot{\phi}\right)=\left(0,0\right)$ 
yields
\begin{eqnarray}
\frac{\bar{r}\cos^{2}\theta}{\left(1-\bar{r}\right)^{2}}
&=& \frac{8\pi  G}{3} \, V_{0} 
\,,\label{eq:poincarestationarypts}\\
\nonumber\\
V_0^{'} &=& 0 \,,\label{eq:poincarestationarypts2}
\end{eqnarray}
where  $\phi_0=\phi(\bar{r}_{0},\:\theta_0)$.  In order to 
satisfy eq. (\ref{eq:poincarestationarypts}) in the
limit $\bar{r}\rightarrow1$ we must have either

\begin{enumerate}

\item $\cos\theta=0$, corresponding to $H\rightarrow0$, 
$\phi\rightarrow\pm\infty$, and $V \left( \phi \rightarrow 
\pm \infty \right)=0 $ (this situation includes potentials 
$V(\phi)$ with compact support).
 
\item $\cos\theta=\pm1$, corresponding to 
$H\rightarrow\pm\infty,\:\phi\rightarrow 0$,
and $V\left( \phi \rightarrow 0 \right) = \infty$ ({\em 
i.e.}, $V$ has a vertical asymptote at $\phi=0$. This is 
the case of the potentials $V(\phi) \propto 
1/\phi^{\alpha}, \alpha>0$ used in many quintessence 
models).

\item $\theta\neq0,\:\pm\pi,\:\pm \pi/2 $, which 
allows $H\rightarrow\pm\infty,\:\phi\rightarrow\pm\infty$,
and $V\left( \phi \rightarrow \pm \infty \right)=\infty$. 
This  case includes unbounded monotonic potentials such as 
$V(\phi)=V_0\, \mbox{e}^{\pm \alpha \phi}$ (scalar field 
cosmology with exponential potentials is studied in 
detail in Ref.~\cite{Wands}).

\end{enumerate}

Fixed points corresponding to any of these situations must 
have a potential that asymptotically satisfies eq. 
(\ref{eq:poincarestationarypts2}) as well as the stated 
conditions. A fixed point satisfying the conditions
of situation~1) corresponds to Minkowski space with no 
potential. Situations~2) and~3) both correspond to extreme 
cases of de Sitter space, with their potentials diverging. 
A possible situation corresponding to case~3) is 
$ V(\phi)=V_{0}\ln\left(\phi/\phi_{0}\right)$.

The next question that one can ask regards the stability 
of these equilibrium points.

\subsection{Stability with respect to homogeneous 
perturbations}

It seems intuitive that if $V(\phi)$ has a local minimum 
at $\phi_{0}$, and a de Sitter equilibrium point 
$\left(H_{0},\phi_{0}\right)$ satisfying
eqs. (\ref{eq:fixdptsH2}) and (\ref{eq:fixdptsV'}) exists, 
it will be stable, and {\em vice-versa}, it will be 
unstable if $V(\phi)$ has a local maximum. However, this 
statement would be a bit naive because $\phi(t)$ couples to 
the other variable $H(t)$ and one must consider 
both variables simultaneously. Here we consider 
homogeneous perturbations of the fixed point 
$\left(H_{0},\phi_{0}\right)$, {\em i.e.}, we write
\begin{eqnarray}
H(t) &=& H_{0}+\epsilon \, \delta_{1}H(t)
+\epsilon^{2}\delta_{2} H(t)+\epsilon^3 
\delta_{3}H(t)+ \, \ldots \,,\label{eq:Hperturbs} \\
&&\nonumber\\
\phi(t) & = & \phi_{0}+\epsilon \, \delta_{1}\phi(t)
+\epsilon^{2}\delta_{2}\phi(t)+\epsilon^{3} 
\delta_{3}\phi(t)+\, \ldots \, \, ,\label{eq:phiperturbs}
\end{eqnarray}
where $\epsilon$ is a smallness parameter and 
$\delta_{(i)}H$ and $ \delta_{(i)}\phi$ depend only on 
time. In 
general, one should consider more general perturbations 
$\delta H(t,\underline{x}),\,\delta\phi (t,\underline{x})$
and even anisotropic perturbations.  Inhomogeneous 
perturbations are subject to notorious gauge dependence 
problems and can only be treated rigorously in the context 
of a gauge-invariant formalism \cite{Bardeen}. This kind of 
formalism is  necessarily very detailed and complicated  
and the corresponding gauge-invariant variables are 
susceptible to physical interpretation only after a gauge 
is fixed (and then, different gauges produce different 
interpretations). For clarity, and to avoid the high degree 
of sophistication needed, we will confine our analysis
to homogeneous perturbations.   
Therefore, if a de Sitter 
fixed point is stable with respect to homogeneous 
perturbations it may still be unstable with respect to 
inhomogeneous ones.   
In the following we assume that an equilibrium point
exists.

By inserting eqs. (\ref{eq:Hperturbs}) and  
(\ref{eq:phiperturbs}) into eqs. 
(\ref{eq:FriedH-dot})--(\ref{eq:FriedKG}) with 
$k=0$ and using the zero order eqs. (\ref{eq:fixdptsH2}) 
and (\ref{eq:fixdptsV'}) for the equilibrium point 
$\left( H_0,\phi_0 \right)$, one obtains the perturbed 
Hamiltonian constraint
\begin{eqnarray}
 &  & 2\epsilon 
H_0\delta_{1}H+2\epsilon^{2}H_{0}\delta_{2}H 
+\epsilon^{2}\delta_{1}H^{2}+2\epsilon^{3} 
H_{0}\delta_{3}H+2\epsilon^{3} \delta_{1}H\delta_{2}H 
\nonumber\\ 
&&\nonumber\\
 &  & =\frac{8\pi G}{3}\left[ 
\frac{1}{2}\left(\epsilon{}^{2} 
\delta_{1}\dot{\phi}^{2}+2\epsilon^{3} 
\delta_{1}  \dot{\phi} 
\, \delta_{2}\dot{\phi}\right)\right.\nonumber\\
&&\nonumber\\
&  &  \left. + \, \frac{1}{2} \, 
V_{0}^{''}\left(\epsilon^{2}\delta_{1}\phi^{2}
+2\epsilon^{3}\delta_{1}\phi\delta_{2}\phi\right)+\, 
\frac{1}{6} \, 
V_{0}^{'''}\left(\epsilon^{3} 
\delta_{1}\phi^{3}\right)\right] 
\,,\label{eq:perturbdH2} 
\end{eqnarray}
the acceleration equation
\begin{eqnarray}
&  & 
\epsilon\delta_{1}\dot{H}+\epsilon^{2}\delta_{2}\dot{H} 
+\epsilon^{3}\delta_{3}\dot{H} = -\left( 2\epsilon 
H_{0}\delta_{1}H+2\epsilon^{2}H_{0}\delta_{2}H 
+2\epsilon^{3}H_{0}\delta_{3}H\right.\nonumber \\
\nonumber\\
 &  & \left.+2\epsilon^{3}\delta_{1}H\delta_{2}H
+\epsilon^{2}\delta_{1}H^{2}\right)
-\frac{8\pi G}{3} \left[ \epsilon^2 
\delta_{1}\dot{\phi}^{2}+2\epsilon^{3}
\delta_{1}\dot{\phi}\delta_{2}\dot{\phi}\right. \nonumber\\
\nonumber\\
& & \left.- \, \frac{1}{2} \, V_{0}^{''} 
\left(\epsilon^{2}\delta_{1}\phi^{2}+2\epsilon^{3} 
\delta_{1}\phi\delta_{2}\phi\right)-\, \frac{1}{6} \, 
V_{0}^{'''}\left(\epsilon^{3}\delta_{1}\phi^{3}\right)\right]  
\,,\label{eq:perturbdHdot}
\end{eqnarray}
and the Klein-Gordon equation
\begin{eqnarray}
 &  & \left(\epsilon\delta_{1}\ddot{\phi} 
+\epsilon^{2}\delta_{2}\ddot{\phi}+\epsilon^{3}
\delta_{3}\ddot{\phi}\right)\nonumber\\
\nonumber\\
 &  & + 3\left(\epsilon 
H_{0}\delta_{1}\dot{\phi}+\epsilon^{2}H_{0} 
\delta_{2}\dot{\phi}+\epsilon^{2}\delta_{1} 
H\delta_{1}\dot{\phi}+\epsilon^{3}H_{0}\delta_{3} 
\dot{\phi}+\epsilon^{3}\delta_{1}H\delta_{2}\dot{\phi} 
+\epsilon^{3}\delta_{2}H\delta_{1}\dot{\phi}\right)\nonumber\\
\nonumber\\
 &  & +V_{0}^{''}\left(\epsilon\delta_{1}\phi+\epsilon^{2} 
\delta_{2}\phi+\epsilon^{3}\delta_{3}\phi\right) 
+\frac{V_{0}^{'''}}{2}\left(\epsilon^{2}\delta_{1}\phi^{2} 
+2\epsilon^3 \delta_{1}\phi 
\, \delta_2 \phi\right)\nonumber\\
&&\nonumber\\
&  & +\frac{V_{0}^{\mathscr{\mathsf{(IV)}}}}{6}  \, 
\epsilon^3\delta_{1} 
\phi^{3}\:=\:0,\label{eq:perturbdKG-1} 
\end{eqnarray}
where $V_{0}^{''}\equiv V^{''}(\phi_{0})$, {\em etc}.  
$V(\phi)$ and $dV/d\phi$ have been expanded to third order 
as
\begin{equation}
V(\phi)=V_{0}+ \frac{V_{0}^{''}}{2} 
\left[\epsilon\delta_{1}\phi(t)+\epsilon^{2} 
\delta_{2}\phi(t)+ \, \ldots \, \right]^{2} 
+\frac{V_{0}^{'''}}{6}\left[\epsilon\delta_{1} 
\phi(t)+ \, \ldots \, \right]^3 +\, \ldots \,, 
\end{equation}

\begin{eqnarray}
 &  & V^{'}(\phi)=V_{0}^{''}\left[\epsilon\delta_{1} 
\phi(t)+\epsilon^{2}\delta_{2}\phi(t)
+\epsilon^{3}\delta_{3}\phi(t)+ \, \ldots \, 
\right]\nonumber\\
&&\nonumber\\
 &  & +\frac{V_{0}^{'''}}{2}\left[\epsilon\delta_{1}\phi(t)
+\epsilon^{2}\delta_{2}\phi(t)+ \, \ldots \, \right]^{2}
+\frac{V_0^{\mathscr{\mathsf{(IV)}}}}{6} 
\left[\epsilon\delta_{1}\phi(t)
+ \, \ldots \, \right]^{3}+ \, \ldots \, ,\nonumber\\
&&
\end{eqnarray}
using the fact that $V_{0}^{'}=0$. To first order in 
$\epsilon$, eq. (\ref{eq:perturbdH2}) yields
$2H_{0}\delta_{1}H=0$ and, if the de Sitter equilibrium 
point is not a degenerate Minkowski space with $H_{0}=0$, 
then\footnote{It is an old adage in 
cosmological perturbation 
theory that there are no linear perturbations of de 
Sitter space sourced by a (minimally coupled) scalar field 
\cite{Grischuk}.} 
\begin{equation}
\delta_{1}H=0 
\end{equation}
and eq. (\ref{eq:perturbdHdot}) then yields 
$\delta_{1}\dot{H}=0$. The Klein-Gordon equation 
(\ref{eq:perturbdKG-1}) then decouples and reduces to the 
equation of the damped harmonic oscillator
\begin{equation}
 \ddot{\delta_{1}\phi+}3H_0 \, \delta_{1}\dot{\phi}
+V_{0}^{''}\delta_{1}\phi=0 \,,
\label{eq:1stperturbKG}
\end{equation}
a second  order ODE with constant coefficients. The
associated algebraic equation is 
\begin{equation}
\lambda^{2}+3H_{0}\lambda+V_{0}^{''}=0 \,,
\end{equation}
which has the roots
\begin{equation}
\lambda_{1,2}=\frac{-3H_{0}\pm\sqrt{9H_{0}^{2} 
-4V_{0}^{''}}}{2}=-\frac{3H_{0}}{2}\left(1\mp 
\sqrt{1-\frac{4V_{0}^{''}}{9 H_0^2}} \, \right) \,.
\end{equation}
If $9H_{0}^{2}-4V_{0}^{''}\neq0$, the general solution of 
eq. (\ref{eq:1stperturbKG}) is 
\begin{eqnarray}
\delta_1 \phi(t) & = & \mbox{e}^{-3H_0 t/2}\left[ C_1 
\exp\left(\frac{\sqrt{9H_0^2-4V_0^{''}} }{2}\:  
t\right) \right.\nonumber\\
&&\nonumber\\
&  + & \left.  C_{2}\exp\left(\frac{-\sqrt{9H_{0}^{2} 
-4V_{0}^{''}}}{2}\: t\right)\right] =C_{1} 
\mbox{e}^{\lambda_{1}t} 
+C_{2} \mbox{e}^{\lambda_{2}t} \,, 
\end{eqnarray}
where $C_{1,2}$ are arbitrary integration constants. It is 
easy to see that $V_{0}^{''}\geq0$ is required for 
stability. In fact, if $V_{0}^{''}\geq0$, then 
$9H_{0}^{2}-4V_{0}^{''}\leq9H_{0}^{2}$. 

If $V_{0}^{''}>9H_{0}^{2}/4=6\pi G \, V_0$, then  
$ \lambda_{1,2}=\left(-3H_{0}\pm i\sqrt{\left|9H_{0}^{2} 
-4V_0^{''}\right|}\right)/2$ have imaginary parts and the 
two independent modes are of the form  
\begin{equation}
\exp\left(\frac{-3H_{0}t}{2}\right) 
\cdot \exp\left(\frac{\pm 
i\sqrt{\left|9H_{0}^{2}-4V_{0}^{''}\right|}}{2}\: t\right) 
\,,
\end{equation}
which decay because of the first exponential if $H_{0}>0$, 
and increase without bound if $H_{0}<0$.

If $V_{0}^{''}=9H_{0}^{2}/4$ then  
$\mbox{\ensuremath{\lambda_{1}=\lambda_{2}=-3H_{0}/2} }$
and the solutions of eq. (\ref{eq:1stperturbKG}) are 
\begin{equation}
\delta_{1}\phi(t)= \mbox{e}^{-3H_0 t/2} \left( C_1+C_2 t 
\right) \,,
\end{equation}
again, stable if $H_{0}>0$.

If $0<V_{0}^{''}<9H_{0}^{2}/4$, then  
$0<9H_{0}^{2}-4V_{0}^{''}<9H_{0}^{2}$,
or $\sqrt{9H_{0}^{2}-4V_{0}^{''}}<3H_{0}$$ $ and 
$\lambda_{1,2}<0$, so that the independent modes 
$e^{\lambda_{1,2}t}$ do not grow.

Finally, if $V_{0}^{''}=0$ in eq. (\ref{eq:1stperturbKG}), 
the solution is 
\begin{equation}
\delta_{1}\phi(t)=C_{1}+C_{2} \, \mbox{e}^{-3H_{0}t} \,;
\end{equation}
this solution is stable for $H_{0}>0$, but not 
asymptotically stable, as it does not fully decay. We can 
therefore conclude that:
\begin{itemize}

\item If $H_{0}>0$, then the de Sitter equilibrium point 
$\left(H_{0},\phi_{0}\right)$ is asymptotically stable if 
$V_{0}^{''}>0$, unstable if $V_{0}^{''}<0 $, and stable but 
not asymptotically stable if $V_{0}=0$.

\item If $H_{0}<0$, the de Sitter equilibrium point 
$\left(H_{0},\phi_{0}\right)$ is always unstable (to first 
order and all higher orders).

\end{itemize}
Furthermore, to first order there is no 
perturbation $\delta_{1}H$ and the 
perturbations $\delta H$ and $\delta\phi$ are decoupled.

The second order equations yield
\begin{eqnarray}
\delta_{2}H &=& \frac{2\pi 
G}{3H_{0}}\left(\delta_{1}\dot{\phi}^{2}
+V_{0}^{''}\delta_{1}\phi^{2}\right) \,,\\
&&\nonumber\\
\delta_{2}\dot{H} & = & -4\pi G \, \delta_{1}\dot{\phi}^{2} 
\,,\\
&&\nonumber\\
\delta_{2}\ddot{\phi} 
&+& 3H_{0}\delta_{2}\dot{\phi}+V_{0}^{''} 
\delta_{2}\phi\,=-\frac{V_{0}^{'''}}{2}\delta_{1}\phi^{2}\, 
.\label{eq:2ndperturbKG}
\end{eqnarray}
The second order perturbations $\delta_{2}H$ and  
$\delta_{2}\phi$ depend on the first order ones 
$\delta_{1}\phi$ and their derivatives. These act as a 
source term in the Klein-Gordon equation 
(\ref{eq:2ndperturbKG}) for $\delta_{2}\phi,$ which is a 
damped forced harmonic oscillator equation. From the 
general theory of ODEs, it is clear that if $H_{0}>0$, the 
friction term $3H_{0}\delta_{2}\dot{\phi}$ will correspond 
to positive friction, while if $H_{0}<0,$ there is 
``anti-friction'' leading to instability.  Therefore,

\begin{itemize}

\item For $H_{0}>0$, there is stability if 
$V_{0}^{''}\geq0$, and instability if $V_{0}^{''}<0$.

\item If $H_{0}<0$, the perturbations $\delta_{2}\phi$ 
grow without bound and the de Sitter equilibrium point 
$\left(H_{0},\phi_{0}\right)$ is unstable.
\end{itemize}

To third order in $\epsilon$, one obtains
\begin{eqnarray}
H_{0}\delta_{3}H &=& \frac{4\pi G}{3} \left( 
\delta_{1}\dot{\phi} \, \delta_{2}\dot{\phi}+V_{0}^{''} 
\delta_{1}\phi \, \delta_{2}\phi 
+ \, \frac{1}{6}\, V_{0}^{'''}\delta_{1}\phi^{3}\right) 
\,,\label{eq:thirdperturbH^2}\\
&&\nonumber\\
\delta_{3}\dot{H} &=& -8\pi 
G \, \delta_{1}\dot{\phi} \, \delta_{2}\dot{\phi} \,, 
\label{thirdperturbH-dot}\\
&&\nonumber\\
\delta_{3}\ddot{\phi} & + & 3H_{0}\delta_{3}\dot{\phi} 
+ V_{0}^{''}\delta_{3}\phi=-\frac{2\pi 
G}{3H_{0}}\left(\delta_{1}\dot{\phi}^{3}+V_{0}^{''}\delta_{1} 
\phi^{2}\delta_{1}\dot{\phi}\right)-V_0^{'''}\delta_1 
\phi\delta_2 \phi 
\nonumber\\
&- & \frac{V_{0}^{\mathscr{\mathsf{(IV)}}}}{6} \, 
\delta_{1}\phi^{3} \,. \label{thirdperturbKG} 
\end{eqnarray}
By considering only situations in which we have stability 
at the lower orders, the lower order perturbations and 
their derivatives, which are acting as sources in the 
higher order equations, are bounded,  and eqs. 
(\ref{eq:thirdperturbH^2}) and (\ref{thirdperturbH-dot})
guarantee that $\delta_{3}H$ and its derivatives are 
bounded by (small) initial conditions. Eq.   
(\ref{thirdperturbKG}) again takes the form of a driven, 
damped harmonic oscillator with  instability if $H_{0}<0$,
or if $H_{0}>0$ with $V_{0}^{''}<0$, and stability for 
$H_{0}>0$ and $V_{0}^{''}\geq0$.

There remains the case of the Minkowski fixed point 
$\left(H_{0},\phi_{0}\right)=\left(0,\phi_{0}\right)$,
a degenerate de Sitter space. This situation occurs if 
$V_{0}=0$ and $V_{0}^{'}=0$. Eqs.  
(\ref{eq:FriedH-dot})--(\ref{eq:FriedKG}) then reduce, to 
third order, to 
\begin{eqnarray}
&& \epsilon^{2}\delta_{1}H^{2}+2\epsilon^{3}\delta_{1} 
H\delta_2 H \nonumber\\
&&\nonumber \\
 & & = \frac{8\pi G}{3}\left\{ 
\frac{1}{2}\left(\epsilon{}^{2}\delta_{1}\dot{\phi}^{2} 
+2\epsilon^{3}\delta_{1}\dot{\phi}\delta_{2}\dot{\phi}\right) 
\right.\nonumber \\ 
&&\nonumber\\
&& \left. +\, \frac{1}{2} 
\, V_{0}^{''}\left(\epsilon^{2}\delta_{1} 
\phi^{2}+2\epsilon^{3}\delta_{1}\phi\delta_{2}\phi\right)+ 
\, \frac{1}{6} \, 
V_{0}^{'''}\left(\epsilon^{3}\delta_{1}\phi^{3}\right)\right\} 
\, , \label{eq:minkowperturbH^2}\\
&&\nonumber\\
&  & \epsilon\delta_{1}\dot{H}+\epsilon^2 \delta_{2}
\dot{H}+\epsilon^{3}\delta_{3}\dot{H}
= -2\epsilon^3 \delta_{1}H\delta_{2}H
-\epsilon^2 \delta_{1}H^{2} \nonumber\\
&&\nonumber\\
&& -\frac{8\pi G}{3}
\left[ \epsilon^{2}\delta_{1}\dot{\phi}^{2}
+2\epsilon^{3}\delta_{1}\dot{\phi} \, \delta_{2}\dot{\phi}
\right.\nonumber \\
&&\nonumber \\
 &  & 
\left.-\frac{1}{2}\, 
V_{0}^{''}\left(\epsilon^{2}\delta_{1}\phi^{2}+2\epsilon^{3}\delta_{1}\phi\delta_{2}\phi\right)-\, 
\frac{1}{6} \, 
V_{0}^{'''}\left(\epsilon^{3}\delta_{1}\phi^{3}\right)\right] 
\,, \label{minkow perturb H-dot}
\end{eqnarray}

\begin{eqnarray}
 &  & \left(\epsilon\delta_{1}\ddot{\phi} 
+\epsilon^{2}\delta_{2}\ddot{\phi}+\epsilon^{3} 
\delta_{3}\ddot{\phi}\right)\nonumber\\ 
&&\nonumber\\
 &  & +3\left(\epsilon^{2}\delta_{1}H\delta_{1}\dot{\phi} 
+\epsilon^{3}\delta_{1}H\delta_{2}\dot{\phi} 
+\epsilon^{3}\delta_{2}H\delta_{1}\dot{\phi}\right)\nonumber\\
&&\nonumber \\
&  & +V_{0}^{''}\left(\epsilon\delta_{1}\phi+\epsilon^{2} 
\delta_{2}\phi+\epsilon^{3}\delta_{3}\phi\right) 
+\frac{V_{0}^{'''}}{2}\left(\epsilon^{2}\delta_{1}\phi^{2} 
+2\epsilon^{3}\delta_1 \phi 
\, \delta_2\phi \right)\nonumber\\
 &  & 
+\frac{V_0^{\mathscr{\mathsf{(IV)}}}}{6} \, 
\epsilon^3 \delta_{1}\phi^{3}=\:0 \,. 
\label{eq:MinkowperturbKG}
\end{eqnarray}
The friction term is now absent in eq. 
(\ref{eq:MinkowperturbKG}). To first order  one obtains 
$\delta_{1}\dot{H}=0$ and 
\begin{equation}
\delta_{1}\ddot{\phi}+V_{0}^{''}\delta_{1}\phi=0 \,,
\end{equation}
therefore there is stability for $V_{0}^{''}\geq0$.

To second order we have
\begin{equation}
\delta_{1}H^{2}=\frac{8\pi G}{3}\left(\frac{\delta_{1} 
\dot{\phi}^{2}}{2}+\frac{V_{0}^{''}\delta_{1}\phi^{2}}{2}\right) 
=\mbox{const.}
\label{eq:minkowperturb2H^2} 
\end{equation}
(which yields no information on second  order 
perturbations),
\begin{equation}
\delta_{2}\dot{H}+\delta_{1}H^{2}=\frac{8\pi 
G}{3}\left(-\delta_{1}\dot{\phi}^{2}
+\, \frac{1}{2} \, 
V_{0}^{''}\delta_{1}\phi^{2}\right)=-4\pi 
G\delta_{1}\dot{\phi}^{2} \, 
,\label{eq:minkowperturb2H-dot} 
\end{equation}

\begin{equation}
\delta_{2}\ddot{\phi}+V_{0}^{''}\delta_{2}\phi
=-3\delta_{1}H\delta_{1}\dot{\phi}-\, \frac{V_{0}^{'''}}{2}
\delta_{1}\phi^{2}  \,.\label{eq:minkowperturb2KG}
\end{equation}
Eq. (\ref{eq:minkowperturb2H-dot}) guarantees stability in 
$\delta_{2}H$,
while eq. (\ref{eq:minkowperturb2KG}) guarantees stability to
second order if $V_{0}^{''}\geq0$.

To third order we have
\begin{equation}
\delta_{2}H=\frac{4\pi G}{3\delta_{1}H}\left(  
\delta_1\dot{\phi} 
\, \delta_2\dot{\phi}+V_{0}^{''}\delta_{1} 
\phi \, \delta_2\phi+\frac{1}{6}\, 
V_{0}^{'''}\delta_{1}\phi^3\right) \,,
\end{equation}
assuming $\delta_{1}H\neq0$ (again, this  equation provides 
no information on third order perturbations),
\begin{equation}
\delta_{3}\dot{H} = - 2\delta_{1}H\delta_{2}H
-\frac{8\pi G}{3} \left( 2\delta_{1}\dot{\phi} \, 
\delta_{2}\dot{\phi}-V_{0}^{''}\delta_{1}\phi\delta_{2}\phi
-\frac{1}{6}\, V_{0}^{'''} \delta_{1}\phi^{3}\right)  
\,,
\end{equation}

\begin{equation}
\delta_{3}\ddot{\phi} 
+V_{0}^{''}\delta_{3}\phi 
=-\left(3\delta_{1}H\delta_{2}\dot{\phi}+3 
\delta_{2}H\delta_{1}\dot{\phi}
+V_0^{'''}\delta_1 \phi\delta_2 \phi
+\frac{V_0^{\mathscr{\mathsf{(IV)}}}}{6} 
\, \delta_{1}\phi^{3}\right) \,,
\end{equation}
which again provides stability for $V_{0}^{''}\geq0$.

\subsection{Inhomogeneous perturbations}

A treatment with respect to inhomogeneous perturbations is 
necessarily more complicated, requiring the use of a 
gauge-invariant formalism (see, {\em e.g.}, 
\cite{Bardeen}). To first order (for which the 
gauge-invariant formalisms apply), the results on the 
stability of de Sitter spaces already obtained also hold  
for inhomogeneous perturbations. In fact, the stability of 
de Sitter spaces in the very general theory of gravity 
described by the action
\begin{equation}
S=\int d^{4}x\sqrt{-g}\left[\frac{f(R,\phi)}{2}
-\frac{\omega(\phi)}{2}g^{\mu\nu}\nabla_{\mu}\phi\nabla_{\nu}\phi
-V(\phi)\right]
\end{equation}
was studied in \cite{key-1}. This class of theories 
contains scalar-tensor gravity if $f(R, \phi)=f(\phi)R$ and 
higher order gravity if $f(R, \phi)=f(R)$ and $\phi=$const. 
A de Sitter space 
$\left(H_{0},\phi_{0}\right)$ is stable with respect to 
inhomogeneous perturbations if and only if \cite{key-1}
\begin{equation}
\left.\frac{\frac{f_{\phi\phi}}{2}- 
V_{\phi\phi}+\frac{6f_{\phi 
R}H^{2}}{f_{R}}}{\omega \left(1+\frac{3f_{\phi 
R}^{2}}{2 \omega f_{R}}\right)}\right|_{\left(H_0, 
\phi_{0}\right)} \leq 0  \,. 
\end{equation}
General relativity with a minimally coupled scalar field 
corresponds to the trivial case
\begin{equation}
f\left(\phi,R\right)=\frac{R}{8\pi G} \,,\,\,\,\;\;\; 
\omega \equiv 1 \,,
\end{equation}
which yields the first order stability  condition
$V_{0}^{''}\geq0$. This result is gauge-invariant and 
reproduces, to first order, the one which we have already 
obtained for homogeneous perturbations (the equivalence 
between homogeneous and inhomogeneous perturbations in 
this respect does not extend to FLRW universes other than 
de Sitter space).

\subsection{Lyapunov functions}

The Lyapunov method \cite{Lyapunov} can be easily applied 
to scalar field  cosmology in order to assess stability 
non-perturbatively,  and to estimate the size of the 
attraction basins of stable  fixed points in the phase 
space.

Let $\left(H_{0},\phi_{0}\right)$ be a fixed point of the 
dynamical system (\ref{eq:FriedH-dot})--(\ref{eq:FriedKG}); 
then, if  $V(\phi)$ has a local minimum at $\phi_{0}$, and 
$H_{0}>0$, the $\mathcal{C}^{1}$ function
\begin{equation}
L_{1}\left(H,\phi,\dot{\phi}\right)
\equiv\frac{\dot{\phi}^{2}}{2}+V(\phi)-V_{0}
\end{equation}
is a Lyapunov function. In fact,

\begin{itemize}

\item $L_{1}\left(H,\phi,\dot{\phi}\right)>0$ in a domain 
$\mathcal{D}$ containing $\phi_{0}$, except at 
$\left(H_{0},\phi_{0}\right)$;

\item $L_{1}\left(H_{0},\phi_{0},0\right)=0$;

\item $dL_{1}/dt=\dot{\phi}\left(\ddot{\phi}+V^{'}\right)
=-3H\dot{\phi}^{2}$ is strictly negative in $\mathcal{D}$, 
except at the fixed point, where $dL_{1}/dt$ vanishes. 

\end{itemize}
Therefore, the fixed point $\left(H_{0},\phi_{0}\right)$ 
is asymptotically stable. If $V(\phi)$ has only one minimum 
at $\phi_{0}$, the attraction basin of 
$\left(H_{0},\phi_{0}\right)$ is the entire phase space, 
while if there are other minima at  
$\phi_{1},\phi_{2}, \, \ldots \, $, the attraction basin of 
$\left(H_{0},\phi_{0}\right)$ is finite and will be limited 
by separatrices between the attraction basins of other
fixed points.

If $H_{0}>0$ and V($\phi)$ has a local maximum at 
$\phi_{0}$ (with $V_{0}^{'}=0,\: V_0^{''}<0$), then the 
$\mathcal{C}^{1}$ function
\begin{equation}
L_{2}\left(H,\phi,\dot{\phi}\right)
=-\frac{\dot{\phi}^{2}}{2}-V(\phi)+V_{0}
\end{equation}
is such that $L_{2}\left(H,\phi,\dot{\phi}\right)<0$ in a 
domain containing $\left(H_{0},\phi_{0}\right)$ at which 
$L_{2}$ vanishes, and $dL_{2}/dt=3H\dot{\phi}^{2}>0$ except 
at the fixed point itself  where $dL_{2}/dt$ vanishes. This 
guarantees that the fixed point is unstable and is a 
repellor.

\section{Late-time behaviour of the solutions}
\setcounter{equation}{0}

Some conclusions on the asymptotic behaviour of the 
solutions at late times can be reached under certain 
assumptions on the scalar field potential, without fully 
specifying the form of $V(\phi)$.

First, for spatially flat universes, eq. 
(\ref{eq:FriedH-dot}) implies that $\dot{H}\leq0$, with the 
equality being satisfied only for the de Sitter fixed 
points. Hence, outside of fixed points, $H(t)$ is always a 
decreasing function. Assuming that $H$ starts out positive  
and that $V(\phi)\geq0$ (which guarantees that the energy 
density is positive), the Hamiltonian constraint 
(\ref{eq:FriedH^2}) shows that $H$ cannot become negative 
because, due to continuity, it would have to vanish first 
and the trajectory of the solution in phase space would 
then cross a fixed point, which is impossible. Therefore,
if $H$ starts out positive {[}negative{]}, it remains 
positive 
{[}negative{]}. Let us consider, for definiteness, the case 
$H>0.$ Since $H$ is bounded from below by zero and 
$\dot{H}<0$, the graph of $H(t)$ cannot cross the $H=0$ 
axis and we must have $\ddot{H}\geq0$ (assuming that 
$H\in \mathcal{C}^{2}$).  Since $\dot{H}(t)=-4\pi 
G \, \dot{\phi}^{2}$, as $\dot{H}(t)\rightarrow0$ at late 
times so also must $\dot{\phi}(t)\rightarrow0$, 
or $\left(H(t),\phi(t)\right)\rightarrow 
\left(H_{0},\phi_{0}\right)$ (with $H_{0}>0$) at late 
times. The trajectories must asymptotically approach a de 
Sitter attractor point. One possibility is that $H_{0}=0$,
corresponding to Minkowski space. This could be a point at 
infinity, in which case the energy content of the universe 
gets diluted in the future expansion and the universe 
more and more resembles  empty Minkowski
space.  In general, for a strictly monotonic 
potential which is non-negative (or otherwise bounded from 
below), we always have $dV/d\phi\neq0$ and there cannot be 
equilibrium points. In this case $ 
\left|\phi\right|\rightarrow+\infty$ at late times with 
$\dot{\phi}\rightarrow0$ due to friction (in an expanding 
universe), and then also $\dot{H}=-4\pi 
G\dot{\phi}^{2}\rightarrow0$. The phase space orbit tends 
to a de Sitter point 
$\left(H_{0},\phi_{0}\right)=\left(8\pi 
GV_{0}/3,\pm\infty\right)$,  where $V_{0}$ is the 
asymptotic value of $V(\phi)$.

A further consequence of the acceleration equation 
$\dot{H}=-4\pi G\dot{\phi}^{2}$ is that there cannot be 
limit cycles as $H$ and $\phi$ would have to repeat 
themselves for periodic orbits, while here $H$ is monotonically
decreasing and thus cannot be periodic.

The late-time asymptotics become much more complicated if 
the scalar field couples non-minimally to the Ricci 
curvature $R$ or if it is a phantom field with the 
``wrong'' 
sign of the kinetic energy \cite{phasespaceNMC, phantomCQG, 
oscillators}. 


\section{Conclusions}
\setcounter{equation}{0}

It is possible to partially analyze the dynamics  
of a  minimally coupled scalar field in FLRW cosmology, 
described by the coupled Friedmann-Klein-Gordon equations 
(\ref{eq:FriedH^2}), (\ref{eq:FriedH-dot}), and 
(\ref{eq:FriedKG}) constituting a  
dynamical  system. It has been shown in the previous 
sections how the study of the geometrical structure of the 
phase space and of the stationary points and their 
attraction/repulsion basins can be carried out without 
specifying the form  of the scalar field potential 
$V(\phi)$. Limiting the  analysis  to spatially flat 
universes reduces the dimension of the phase space from 
four to three. Further, the trajectories of the 
system lie  on two intersecting energy ``sheets'' 
(corresponding to  setting an effective  Hamiltonian equal 
to zero) given by the Hamiltonian
constraint (\ref{eq:phasespaceconstraint}). The two 
sheets correspond to $\dot{\phi}>0$ and $\dot{\phi}<0$. 
There is no  possibility of chaos in this space due to its 
lower dimensionality.  The dynamics of non-spatially flat 
universes  are described by orbits constrained to take 
place above the upper sheet and  below the lower sheet  
$\left( |\dot{\phi}| >|\dot{\phi}_{flat}|\right)$
for $k> 0$, and between these sheets for $k<0$.

Any stationary points are necessarily de  Sitter
universes with a constant scalar field, which includes  
Minkowski space as the special case $H_0=0$. These 
points were found to be stable for $H_{0}\geq0$ and 
$V_{0}^{''}\geq0$, by both third order 
homogeneous perturbation  analysis and
Lyapunov's second method. The size of their attraction 
basins is  dictated by the form of the potential. The 
attraction  basins are global if  $V_{0}^{'}=0$  
corresponds to a  single sink. The possibility of 
limit cycles is excluded.

In an asymptotic analysis,  if we assume that $H$ 
starts off positive (an initially expanding universe) and 
$V(\phi)>0$ (enforcing the weak energy condition), then  
$H(t)$ will remain positive for all times and 
all  trajectories meeting these conditions  will be 
asymptotically attracted to  de Sitter spaces. 
Physically, this means that for these conditions 
the universe will always  expand, with the scale factor 
$a(t)$  becoming exponential at late times. Stationary 
points could exist as an  infinite limit in 
$H$ and/or $\phi$ for certain potentials.

From the previous results one can see that 
an FLRW  universe containing a single scalar field in 
general relativity has  relatively simple dynamics. The 
rather straightforward discussion presented here needs, of 
course, to  be supplemented by a more detailed study which 
can only  be carried out by fully specifying the potential 
$V(\phi )$, and this is the limitation of the present 
paper. 
At the same time, not specifying the potential makes our 
discussion completely general. An added bonus of the 
material presented here is its pedagogical value for a 
general introduction to inflation and quintessence models.

\section*{Acknowledgments}

We thank Andres Zambrano for discussions and Hugues 
Beauchesne for help with one of the 
figures.  This work is supported by the Natural Sciences 
and Engineering Research Council of Canada (NSERC).


\end{document}